\documentclass[]{mn2e}
\title{A Circular Planetary Nebula around the OH/IR Star OH\,354.88-0.54  (V1018 Sco)}
\author [Martin Cohen et al.]
{Martin Cohen$^{1}$, Quentin A. Parker$^{2,3}$, Jessica Chapman$^{4}$ \\
$^{1}$Radio Astronomy Laboratory, University of California,
    Berkeley, CA 94720\\
$^{2}$Department of Physics, Macquarie University, Sydney, NSW 2109 Australia\\
$^{3}$Anglo-Australian Observatory, PO Box 296, Epping, NSW 2121, Australia\\
$^{4}$Australia Telescope National Facility, PO Box 76,  Epping, NSW 2121, Australia}

\date{ Accepted . Received ; in original form        }

\begin{document}

\maketitle

\begin{abstract}
New deep, high-resolution H$\alpha$ imagery from the UK Schmidt Telescope (UKST) Unit's 
H$\alpha$ survey of the Southern Galactic Plane reveals the presence of a faint, highly 
circular, planetary nebula surrounding a very long period variable star (now known as V1018 
Sco), first discovered as a 1612-MHz OH maser, OH\,354.88-0.54.  The nebular phase-lag distance,
diameter, and radial velocity are 3.2~kpc, 0.3~pc, and 13~km~s$^{-1}$, respectively.  Combining the
maser attributes with near-, mid-, and far-infrared data and with our optical spectrum of 
the ring we conclude that the object was an intermediate mass AGB star 
(initial stellar mass $\geq4M_{\odot}$) in which the fast wind has recently turned on, ionizing previously shed circumstellar
material.  Hence, we speculate that we may be witnessing a hitherto unobserved phase of PN evolution, 
in which a PN has only recently started to form around a star that is unequivocally still in its AGB
phase.
\end{abstract}


\section{INTRODUCTION}
The source,``AFGL~5356'', represents the first recognition of anything unusual at 
{\it l}=354.88$^\circ$, {\it b}=-0.54$^\circ$ in the form of the discovery of mid-infrared (MIR) 
emission from a point source in the RAFGL Rocket Sky Survey of Price \& Walker (1976). 
OH\,354.88-0.54 was 
independently discovered as a strong 1612-MHz maser during a maser survey of the 
Galactic plane, between {\it l}=340$^\circ$ and the Galactic Centre, by Caswell et al. (1981).
The 1612-MHz spectrum showed two well-defined peaks with an expansion velocity 
of $\sim$15\,km\,s$^{-1}$, typical for asymptotic giant branch (AGB) stars with high mass-loss 
rates. By the early 1980s, OH\,354.88-0.54 was recognized to be an extremely heavily obscured star
with no detectable optical counterpart. The next 15 years saw a 
variety of infrared (IR) measurements of the source, culminating in a secure identification of the object 
as an extreme, large amplitude, long period variable, with several unusual characteristics.

In this paper we bring together the key multi-wavelength data on OH\,354.88-0.54.  
We introduce the nebular ring (\S2), summarize what is currently known about 
OH\,354.88-0.54 from the perspective of the maser (\S3),
the IR photometry (\S4), and the optical spectrum of 
the nebula (\S5), leading to our conclusions about the nature of the nebula (\S6).

To provide a framework within which to locate the elements of OH\,354.88-0.54 described in this paper, 
we offer the following cross-section of a typical AGB star's circumstellar shell (Reid \& Menten 
(1997: their Figure~12), Cohen (1989), and Chapman \& Cohen (1986)) by radius and temperature.
Stellar photosphere: $3\times10^{13}$~cm, 2400~K; SiO masers: $8\times10^{13}$~cm, 1250~K; dust 
formation zone: $10^{14}$~cm, 1100~K; OH 1612-MHz masers: $1\times10^{16}-1\times10^{17}$~cm, 
$\sim$50~K.

The unexpected optical detection of the H$\alpha$ ring nebula clearly associated with the OH/IR
star has prompted this work because of what it might tell us about the earliest onset of the fast
wind in PN formation.  The overriding
issues are to identify the source of ionization for this nebula, to explain its 
visibility and regular form at a time when the central star is still a
long period variable, and hence locate
it in the context of late stellar evolution as it proceeds to the PN phase.

\section{DISCOVERY OF THE SURROUNDING H$\alpha$ EMISSION NEBULA}
\begin{figure} 
\vspace{7cm} 
\includegraphics{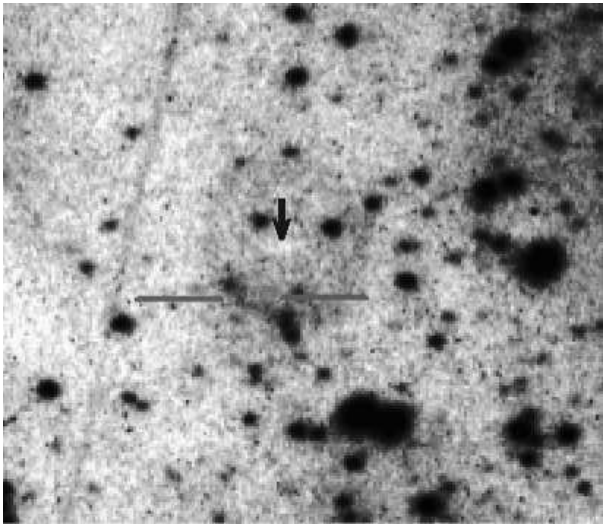} 
\caption{The delicate circular H$\alpha$ ring surrounding the position of
OH\,354.88-0.54, plotted in Galactic coordinates.  The two lines indicate
the location of the slit during follow-on spectroscopy (see \S5). The gap between the lines
is to show where we observed the brightest nebulosity around the ring.
The optically invisible star lies at the head of the arrow.
North is to the top and east to the left.  The image size is 128$^{\prime\prime}$ in Dec.
by 140$^{\prime\prime}$ in RA.}
\label{ring}
\end{figure}

The region of the sky where OH\,354.88-0.54 is located shows much complex and diffuse nebular emission.
The AAO/UKST H$\alpha$ survey of the Southern Milky Way (Parker \& Phillipps 1998; Parker \& 
Phillipps 2003) took a survey exposure of field HA630 (exposure number HA18520) on which  
a distinct, circularly symmetric nebula (Figure~\ref{ring}) was discovered.  
This was during a systematic search for Galactic Planetary Nebulae (PNe) derived from 
this survey (Parker et al. 2003a).  Version 1.0 of the
Edinburgh/AAO/Strasbourg Catalogue of Galactic Planetary Nebulae has been released
(Parker et al. 2003b), and the effort continues as
the Macquarie-AAO-Strasbourg H$\alpha$ PN project (hereafter referred to as the ``MASH").  
The object appears projected against the edge of the highly opaque dark cloud G354.9-00.6 
(Hartley et al. 1986), measuring about $4^{\prime}\times3^{\prime}$.  
The size of the nebula was measured by overlaying circles with a wide variety of 
centres and radii from which
both the circularity and 39\,$\pm$1$^{\prime\prime}$ diameter were determined.
As part of the standard procedure for comparing detections of new PN candidates with
possible alternative identifications, it was noted that the position of the centroid of the
ring (17$^h$ 35$^m$ 02.62$^s$$\pm$0.16$^s$, 
$-$33$^\circ$ 33$^\prime$ 27.92$^{\prime\prime}$$\pm$1.90$^{\prime\prime}$, J2000)
very closely matched the position of a known strong OH maser source, OH\,354.88-0.54.
(The positional uncertainties arise from the difficulty in defining the outer edges of the faint ring.)
Details of the nebula are given in Table~\ref{bdata}, including data presented later in this paper.

\begin{table}
\begin{center}
\caption{Details of the Planetary Nebula around OH\,354.88-0.54}
\label{bdata}
\begin{tabular}{lc} 
\hline
{\it MASH}& \\
RA (J2000)             & $17^{\rmn h}~35^{\rmn m}~02.62^{\rmn s}$ \\
Dec (J2000)            & $-33^\circ~33'~27.9''$                   \\
l, b                   & 354.88$^\circ$, $-$0.54$^\circ$            \\
Diameter (outer)       & 39\,arcsec                               \\
H$\alpha$ Survey Film  & HA18520                                  \\
H$\alpha$ Survey Field & 630                                      \\
Date of H$\alpha$ image & 1999 August 6      \\
PN designation         & PN G354.8$-$0.5                            \\
PHR PN designation     & PHR1735$-$3333                          \\
{\it Literature}       & \\
AFGL~5356 RA(J2000)    & $17^{\rmn h}~35^{\rmn m}~01.8^{\rmn s}$ \\
AFGL~5356 DEC(J2000)   & $-33^\circ~33'~31''$                   \\
{\it IRAS}17317$-$3331 RA(2000)& $17^{\rmn h}~35^{\rmn m}~02.2^{\rmn s}$ \\
{\it IRAS}17317$-$3331 DEC(2000)& $-33^\circ~33'~30''$                   \\
{\it This work}& \\
Average NIR/MIR RA(2000)       & $17^{\rmn h}~35^{\rmn m}~02.72^{\rmn s}$\\
Average NIR/MIR DEC(2000)      & $-33^\circ~33'~29.40''$                   \\
Average OH maser RA(2000)      & $17^{\rmn h}~35^{\rmn m}~02.73^{\rmn s}$ \\
Average OH maser DEC(2000)     &$-33^\circ~33'~29.41''$                   \\
Distance               & 3.2~kpc  \\
Diameter               & 0.3~pc \\
Radial velocity  (LSR) & -10~km~s$^{-1}$ \\
Density                & 4000~cm$^{-3}$ \\
Stellar velocity (LSR) & 9.4~km~s$^{-1}$ \\
Ionized gas mass       & 0.35~M$_\odot$\\
Filling factor         & 0.02\\
\hline
\end{tabular}
\end{center}
\end{table}

Our ongoing multi-wavelength studies of new MASH PNe (e.g. Cohen \& Parker 2003)
includes cross-checking against imagery from the MSX MIR mission.
This led us to recognize a very bright MIR source at the centre of the H$\alpha$
ring.  The astrometry of MSX images led to its 
identification with the known OH maser and with the {\it IRAS} source, 17317-3331.  Note that this
star is so bright in the MIR that the brightest pixels create artifacts in the images that result in
an atypically poor astrometric position for MSX, some 3.7$^{\prime\prime}$ away from the 
best estimate of the OH/IR star's position (see \S3.3).

\section{The masers}
\subsection{OH 1612-MHz spectral characteristics}
\begin{figure} 
\vspace{6.5cm}  
\includegraphics{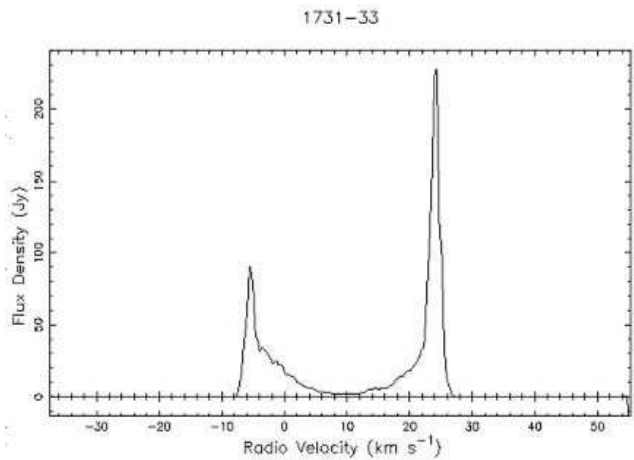}
\caption{The average 1612-MHz spectral profile of OH\,354.88-0.54  from 36 separate epochs
of observation at Parkes.  Note the non-zero flux level at the centre of the profile, near
the stellar velocity.  All velocities are reduced to LSR values.}
\label{jessoh}
\end{figure}

OH monitoring of OH/IR stars was carried out by several groups during the 1980s (e.g.
Herman \& Habing 1985, van Langervelde et al. 1990).  A long-running programme at the ATNF of
monitoring OH/IR masers associated with {\it IRAS} sources, by
Chapman and colleagues, observed OH\,354.88-0.54 using Parkes at 1612~MHz during 36 epochs between 
1988 December and 1993 July.  It is seen to have the classic double-peaked spectrum of a strong OH/IR
star, but its strength is such that emission is detected across the stellar velocity, rather 
than falling to zero, as is more typical of this class of maser.
Figure~\ref{jessoh} presents the averaged profile from these 36 data sets; these data have never
been previously published.  The blue-shifted 
peak velocity is -5.5 km s$^{-1}$ and the red-shifted peak velocity is +24.3 km s$^{-1}$.  These
indicate that the systemic velocity of the star is +9.4 km s$^{-1}$ (LSR) and the shell is
expanding at 14.9 km s$^{-1}$, where each velocity is measured to a precision of 0.18 km s$^{-1}$.
This systemic velocity does not correspond to a plausible distance for the star based on 
Galactic rotation (24~kpc), indicative of a peculiar velocity.

Frail \& Beasley (1994) sought OH/IR stars in a sample of globular clusters, noting the proximity in 
the sky of OH\,354.88-0.5 to the cluster, Liller~1.  They rejected any association with the cluster on
the basis of radial velocity.
Sevenster et al. (1997a) reobserved the maser during a systematic survey for sources of 
OH 1612 MHz maser emission in the Galactic plane and Bulge in late 1993.  The 
spectral profile derived from her data is entirely consistent with that in Figure~\ref{jessoh}, 
although the flux densities of both peaks are lower than in our data because Sevenster et al. 
observed near IR
minimum light.  The last 1612-MHz observation of which we are aware was taken on 1999 November 11 
at the Australia Telescope Compact Array while setting up for an OH observing run.  
The spectral profile indicates that OH\,354.88-0.54's OH emission character had not 
changed at a time only a few months after the H$\alpha$ exposure was taken.

\subsection{Light curve, stellar period, and phase lag}
\begin{figure} 
\vspace{7cm} 
\includegraphics{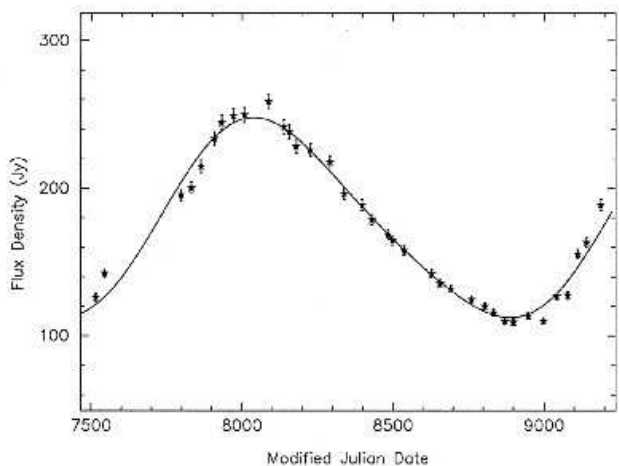}
\caption{The 1612-MHz light curve of OH\,354.88-0.54  assembled from the 36 epochs of 
Parkes' data, from a weighted average of the emission across the blue and red-shifted
peaks.  The best-fit curve for a period of 1486~days is shown fitted to these data.  
Dates are JD expressed as 2440000+the abscissa.}
\label{ohltcrv}
\end{figure}

The light curve for the maser (expressed as the average of the front and back peaks), 
corresponding to the data averaged in Figure~\ref{jessoh}, is shown in Figure~\ref{ohltcrv}, 
from which the 1612-MHz period is determined to be 1486$\pm$20 days (Chapman et al. 1995).  The 
associated best fitting curve also appears in this figure, on the basis of a simple 
combination of the fundamental period and a single harmonic.

\begin{figure} 
\vspace{7.5cm} 
\includegraphics{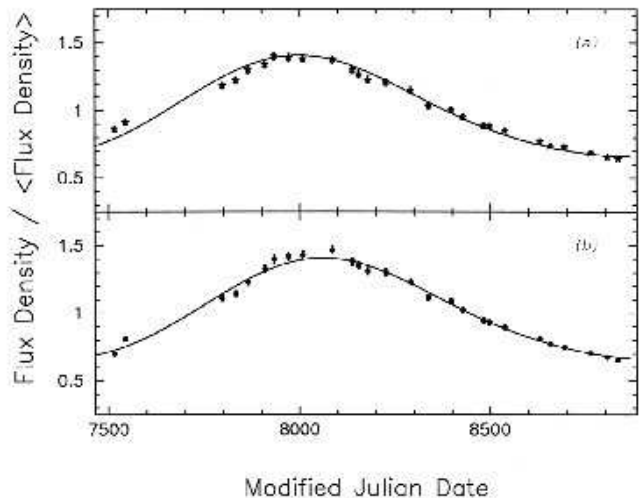}
\caption{Light curves for the OH 1612-MHz blue-shifted ((a) top) and red-shifted
((b) bottom) emission peaks of OH\,354.88-0.54, corresponding to velocities of
-5.5 and +24.3 km~s$^{-1}$, respectively. In each case, the peak flux densities
are normalized by the mean flux density. The phase lag between the front
and back of the circumstellar envelope is clearly evident.  Each curve is shown
using our best-fit period.}
\label{lag}
\end{figure}

\noindent
The linear size of a circumstellar envelope can be determined by
measuring the phase lag, or time delay between the arrival times of
maser emission from the front and back of the circumstellar
envelope. For OH\,354.88-0.54, the measured phase lag is one of the
largest known among the 95 sources monitored at Parkes.  The best-fit
value has been determined by separately fitting a 1486-day period to blue and red peaks to 
assess the arrival time of maxima.  The lag is 60$\pm$9 days, and corresponds to a linear 
diameter of $1.6\times10^{17}$ cm (11000~AU), making this an unusually large shell. The lag
can readily be seen in Figure~\ref{lag} which represents the first appearance 
of these data and their fits.  While this is not the largest phase lag known for such a maser,
one can gain a sense of just how rare such large phase lags are from the analysis by van Langevelde
et al. (1990) of a sample of 1612-MHz masers monitored at Dwingeloo.  Of 43 masers with definite
OH variability, only three have significantly detected lags that exceed the 60 days of OH\,354.88-0.54,
and all three have very long periods comparable to or greater than that of OH\,354.88-0.54.

\subsection{Maser position refinement}
Five independent sets of OH maser positions are available.
It is important to establish that these represent the same object - the long period variable - rather
than a set of ``hot spots" that appear and disappear over  time like those associated with
cool supergiant sources like VY CMa.  These maser positions are due to:
Bowers \& Knapp (1989) with the Very Large Array (VLA) in A/B-configuration at 1612~MHz;
Becker, White \& Proctor (1992) with the VLA in B-array at 1612~MHz;
Fix \& Mutel (1984) with the VLA in A-configuration at 1612~MHz; Sevenster et al. (1997a,b) 
with the Australia Telescope Compact Array (ATCA) at 1612~MHz; and Caswell (1998), 
through high signal-to-noise observations of the associated 1665/1667-MHz emission.
The ratio of peak emissions of the maser at 1665:1667:1612 MHz is roughly 1:4:400,
and these are typical ratios for OH/IR stars. 
The VLA has a significantly smaller beam size than the ATCA at 1612~MHz, but the ATCA positions have
smaller formal uncertainties so we are
inclined simply to average all five positions with equal weights.  This yields a best
estimate for the position of the star itself as: 17$^h$ 35$^m$ 02.73$^s$$\pm0.02^s$,  
-33$^\circ$ 33$^\prime$ 29.41$^{\prime\prime}$$\pm0.24^{\prime\prime}$.  Therefore, the estimated
centre of the H$\alpha$ ring lies within 1$\sigma$ of the stellar position.

\subsection{OH 1612 MHz maser distribution and stellar distance} 
Welty et al. (1987) published well-resolved, VLA, A-array images of the
OH 1612-MHz shell of OH354.88-0.54. The emission was
clearly strongest in an E-W (east-west) plane and weaker to the N (north) and S (south).
We interpret this as a well-filled torus around the star with less
dense emission along a polar axis. From their maps near the stellar velocity we 
estimate the angular diameter to be 3.3$\pm$0.3$^{\prime\prime}$, 
measured from the separation of the prominent pair of emission knots.

Combining the phase lag diameter of $1.6\times10^{17}$ cm with this angular diameter
yields a distance of 3.2~kpc. The likely uncertainty on this distance estimate is about
20\% from the combination in quadrature of a 15\% error in phase lag and 15\% in
angular diameter.  While the phase lag method must yield the most accurate distance estimate,
it is noteworthy that Le Sidaner \& Le Bertre (1996) cite a period-luminosity distance of 3.5~kpc,
although they finally adopted 3~kpc in their modeling of this star.  

\subsection{Other masers in OH\,354.88-0.54}
86-GHz SiO maser emission is very weak in this star, but definitely present.  Nyman et al. (1993) found 
unusual SiO properties for OH\,354.88-0.54 with a ratio of integrated
43-GHz ({\it v}=1, J=1$\rightarrow$0) to 86-GHz ({\it v}=1, J=2$\rightarrow$1) emission
$>$18, and a very high ratio of the J=1$\rightarrow$0 $v=2$/$v=1$ 
transitions at 43 GHz of 4.4.  The average 43/86-GHz line strength ratio is only 0.42 in Miras
and supergiants (Lane 1982), while the 43-GHz line ratio is 
$\sim$1 in Miras, supergiants and semi-regular variables (e.g. Lane 1982) and
$\sim$2 in warmer OH/IR stars (Nyman et al. 1993).  In their survey of 313 inner Galaxy {\it IRAS} 
sources with the colours of late-type stars, Jiang et al. (1995) found 
no object with an integrated 43-GHz line ratio greater than 4.  
Nakashima \& Deguchi (2003) remeasured the 43-GHz SiO maser lines in 
OH\,354.88-0.54, finding a ratio of 3.7, less extreme than that of 
Nyman et al. (1993).  Nakashima \& Deguchi (2003) also present a valuable diagram correlating this 
line ratio in many {\it IRAS} selected masers with [12-25]
colour index.  From this diagram one can see that OH\,354.88-0.54 is by no means unique in terms of
this line ratio; 7\% of the sample of 143 masers have ratios above 4.

\subsection{The nature of OH\,354.88-0.54 }
The OH 1612-Hz maser data are characteristic of maser emission from an
OH/IR star with a high expansion velocity ($\sim$15 km s$^{-1}$), an 
exceptionally long pulsation period (1500~days), and an extremely large OH shell
($1.6\times10^{17}$ cm). The large-amplitude pulsations evident in the OH light
curves indicate that the star is still pulsating strongly, and has not
yet left the AGB stage of evolution, while the large
masing envelope size and high expansion velocity indicate a fairly high mass
star with an initial stellar mass of $\geq4M_\odot$ (and probably more: 
see Chapman et al. 1995).  Neither irregular nor bipolar outflows are inferred from 
the OH maser spectral profile or spatial distribution.

\section{Infrared observations}
\subsection{Near-IR measurements} 
Jones et al. (1982) sought the IR counterpart of OH\,354.88-0.54 in $L^\prime$, while 
Epchtein et al. (1982) searched in $K$.  This difference in strategy
most likely accounts for the discrepancies of 4$^m$ in $K$ and $L^\prime$
between the contemporaneous photometry from these two groups.  The real source 
was very faint in $K$ during early 1981, and probably undetected by Epchtein et al., 
causing them to identify a different object which they characterized as a probable reddened 
field star rather than the true IR counterpart of OH\,354.88-0.54.
Le Bertre (1988) estimated its period from near-infrared (NIR) monitoring to be $>$1200 days.  
Le Bertre \& Nyman (1990) reported OH\,354.88-0.54  to be one of the reddest known OH/IR objects, 
with a $K-L$ colour index $>$6.  Long-term IR monitoring at 15 epochs between March 1986 and July 
1990 in the $L^\prime$ and $M$ bands enabled Le Bertre (1993) to derive periods of 1418 ($L^\prime$) 
and 1448 days ($M$).  The average of these IR periods, 
1433 days, was cited by Nyman et al. (1993), and is consistent with our OH-based period at the 
1$\sigma$ level.  This long period and large amplitude are entirely in keeping with
the NIR redness and faintness at $J$, below 16th magnitude when observed by 2MASS 
(on JD2451039 in August 1998, at phase 0.14, close to the expected epoch of another 
IR minimum).  Inspection of the 2MASS trio of images corroborates the $J$ and $H$
invisibility of the star, even at levels well below those of the 2MASS point source archive, 
but readily reveals the object in the $K$ band.  The minimum amplitudes of NIR 
variability, based on combining 2MASS data with those in the literature, are: 
$\Delta$$J$~$\sim4.9$, $\Delta$$H$~$\sim5.4$, $\Delta$$K$~$\sim3.4$, 
$\Delta$$L^{\prime}$~$\sim6.9$,  $\Delta$$M$~$\sim6.3$. OH\,354.88-0.54 is indeed an extreme variable.  

\subsection{Mid-IR measurements}
The MIR emission of OH\,354.88-0.54 was first measured at 4, 11, and 20~$\mu$m (Price \& Walker 1976).
{\it IRAS} subsequently detected broadband emission at 12, 25, and 60~$\mu$m. 
 OH\,354.88-0.54  was imaged using the IR speckle 
technique at 10~$\mu$m by Cobb \& Fix (1987) who drew attention to the unusual degree of asymmetry 
between its measured N-S and E-W diameters [0.37$\pm$0.015$^{\prime\prime}$, 
0.23$\pm$0.025$^{\prime\prime}$, with the ratio of N-S/E-W of 1.63$\pm$0.19]. At the distance of
3.2~kpc, these translate into physical dimensions of 1.8$\times$10$^{16}$ (N-S) and 
1.1$\times$10$^{16}$~cm (E-W), roughly a factor of 10 smaller than the E-W extent of the 1612-MHz
distribution.

\begin{figure} 
\vspace{7.5cm}  
\includegraphics{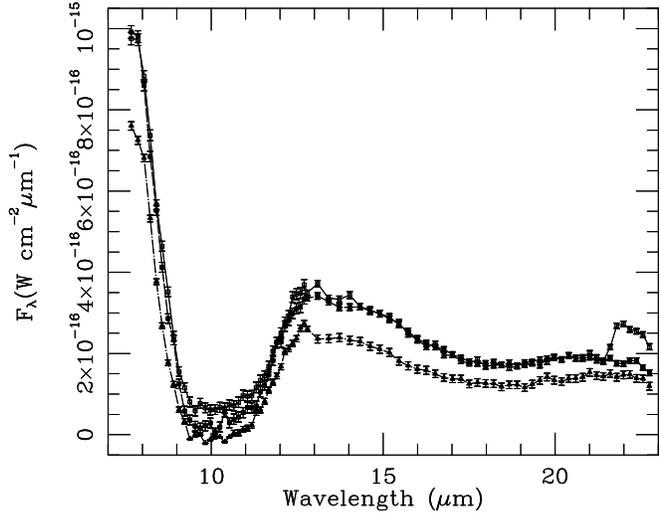}
\caption{The three independent {\it IRAS} LRS spectra of OH\,354.88-0.54 from different
phases of the MIR light curve.  1$\sigma$ error bars are shown on all points.
The apparent feature near 22~$\mu$m in the upper spectrum is caused by an electronic 
glitch in the red spectrometer's baseline during that spectrum particular.}
\label{lrs}
\end{figure}

We have extracted three individual IRAS spectra of this source from the Low Resolution Spectrometer 
(LRS) database
and have corrected their shapes and recalibrated them as described by Cohen et al. (1992a,b), using
the contemporaneous 12-$\mu$m measurements from the IRAS Working Survey Data Base (WSDB). 
Figure~\ref{lrs} illustrates these time-resolved spectra.  Circumstellar silicate 
absorption features show deeply at 10~$\mu$m, with the associated broad, shallow 
absorptions centred near 18~$\mu$m.  The time separations between the two (almost 
coincident) upper LRS spectra, and between the second and the third spectrum were 7 and 166 days, 
respectively, spanning a total interval of 0.11 in light curve phase.  The optical depths in the 
10~$\mu$m silicate feature ($\tau_{10}$) were 2.3, 3.0, and 4.0, respectively, and are somewhat 
dependent on the interpolated continua.  This shows the pattern of increasing $\tau_{10}$ as
the star declines toward minimum light as predicted by Le Bertre (1993) for AGB stars with very
thick dust shells.  These values suggest a total line-of-sight extinction of A$_V$~$\sim$~40-80 
magnitudes.

Le Sidaner \& Le Bertre (1996; their Table 4, final column) attribute 1.7~mag at 10~$\mu$m 
to the purely interstellar extinction toward this star.  Using the plots of $E(B-V)$ with distance by
Fitzgerald (1968) and Lucke (1978) we estimate from our phase-lag distance that the star must suffer
an A$_V$ from 1-3.5 mag.  Therefore, the circumstellar reddening dominates the interstellar component.  

These LRS spectra are very typical of the most extremely red O-rich LPVs (see Fig.~2a of
Wainscoat et al. 1992).  The star has been modeled by Suh (1999: using the mean LRS spectrum), who 
adopts $\tau_{10}$=20, a stellar temperature of 2000~K and luminosity of 10$^4$~L$_\odot$, and a dust opacity 
function specific to cool silicate grains to fit both silicate absorption features. Le Sidaner \& Le Bertre 
(1996: using IR photometry) use 1800~K, 1.7$\times$10$^4$~L$_\odot$, and $\tau_{10}$=11.2.  They derive a 
substantial dust mass loss rate of 7$\times$10$^{-7}$~M$_\odot$~yr$^{-1}$.  Using their
value of 100 for the circumstellar gas-to-dust ratio, this gives a total mass loss rate of
7$\times$10$^{-5}$~M$_\odot$~yr$^{-1}$, consistent with a star of at least intermediate mass.
Chen et al. (2001) argue from their analysis of 1024 OH maser sources with
LRS spectra that those with silicate absorption represent a higher mass population than the more
frequently encountered AGB stars with silicate emission. 

The MSX Galactic plane survey and its Point Source Catalog (PSC1.2) show a very bright MIR source 
(Table~2) at the OH maser's location.
Using the embedded $2-35~\mu$m spectral library from the SKY model of the point source sky (Cohen 1994) 
we have synthesized the colours of all 87 categories of IR source in SKY from the relative response curves 
for the MSX bands. On the basis of comparison with these 87 classes of object we determine that 
OH\,354.88-0.54 is a star that is slightly more extreme than the reddest O-rich AGBM stars 
represented in the SKY model. 

\subsection{The IR position of the star}
The weighted average location of the combination of 2MASS $K$, Jones et al. (1982) $L^\prime$, {\it IRAS} 
and MSX positions is 17$^h$ 35$^m$ 02.72$^s$$\pm$0.13$^s$, 
$-$33$^\circ$ 33$^\prime$ 29.40$^{\prime\prime}$$\pm$1.83$^{\prime\prime}$, which is in excellent agreement
(well within 1$\sigma$) with the accurate OH position for the star and with the estimated centre of the
H$\alpha$ ring.

\subsection{Mid- and far-IR colour-colour planes}
On the basis of its {\it IRAS} colours, OH\,354.88-0.54 was variously described as a young 
stellar object (of T Tauri type) in a dark cloud 
(Persi et al. 1990), a young PN (Zijlstra et al. 1989), and was 
rediscovered as an {\it IRAS} source having the colours of a PN in the 12-25/25-60~$\mu$m
plane by Garcia-Lario et al. (1997), who also made NIR photometric measurements
between 1 and 5~$\mu$m.  This varied character arises because these colour-colour boxes or 
``occupation zones'' often overlap (Walker \& Cohen 1988; Walker et al. 1989), despite the convenience 
of representing colour-colour planes as having distinct compartments (van der Veen \& Habing 1988).

The IRAS [12-25],[25-60] colour-colour plane has been widely utilized over the past two decades as 
a tool for diagnosing the nature of {\it IRAS} sources (e.g., Walker \& Cohen 1988; van der Veen \& Habing 
1988).  Its specific application to OH/IR stars has been pursued most recently by Sevenster et al. 
(1997a,b) and Sevenster (2002a,b) using colours [12-25]~=~2.5log$_{10}$F$_\nu$(25)/F$_\nu$(12) and
             [25-60]~=~2.5log$_{10}$F$_\nu$(60)/F$_\nu$(25).  All flux densities are measured in Jy.
Sevenster's IRAS colours, like her MSX colours, follow the convention that excludes the
zero point constants, offsetting her planes from any absolutely-defined versions of these.  To avoid
confusion we offer only diagrams based on this popular convention in this paper.

Sevenster (2002b) has explored the separation between AGB and post-AGB stars, dividing OH/IR stars
into ``RI'' and ``LI'' groups. The distinction between these two groups is intended to be whether 
their location in this {\it IRAS} colour-colour diagram is to the left (LI) or right (RI) of the
evolutionary sequence first defined in this plane by van de Veen \& Habing (1988).
Miras and OH/IR stars are located on or very near this evolutionary track.  Most well-studied 
post-AGB stars are in the RI region.  This is the traditional region for finding post-AGB stars
and Sevenster (2002b) has argued that these are lower-mass stars ($\sim$2M$_\odot$) that will become
elliptical/circular PNe.  The LI region, discussed extensively only by Sevenster (2002a,b),
is associated with intermediate-mass AGB stars ($\geq4M_\odot$) which Sevenster speculates are 
more likely to become bipolar PNe.  This is also in accord with the much lower mean Galactic latitude
of the LI sources (0.46$^\circ$) compared with that for RI sources (1.66$^\circ$). The average shell 
expansion velocities of OH/IR stars in the RI group are below 15~km s$^{-1}$, while those of the LI group
are above (Sevenster 2002a,b).  

\begin{figure} 
\vspace{7.3cm} 
\includegraphics{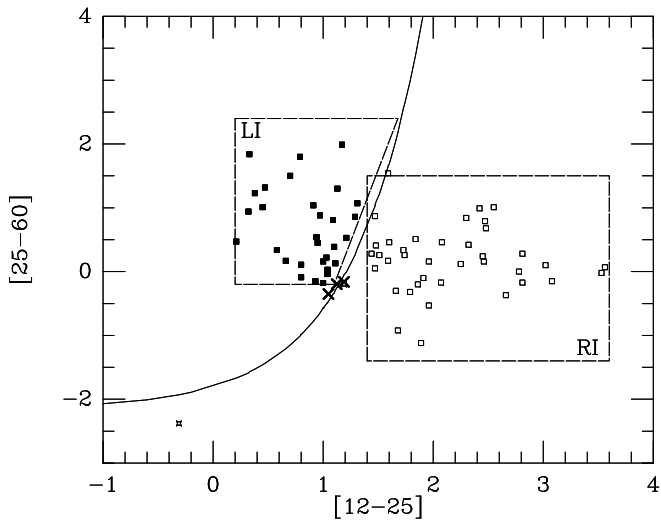}
\caption{{\it IRAS} colour-colour plane showing an attempt to separate OH maser sources into LI and RI
zones (the dashed boxes), and comparing these with the suggested evolutionary sequence of
van de Veen \& Habing (1988). OH\,354.88-0.54's  three independent {\it IRAS} observations are represented 
by the set of three large crosses along this sequence near the lower right corner of the LI 
zone, moving in time up and to the right along the track. The symbols denote LI (filled squares) 
and RI (open squares) stars.}
\label{msiras}
\end{figure}

\begin{figure} 
\vspace{7.5cm} 
\includegraphics{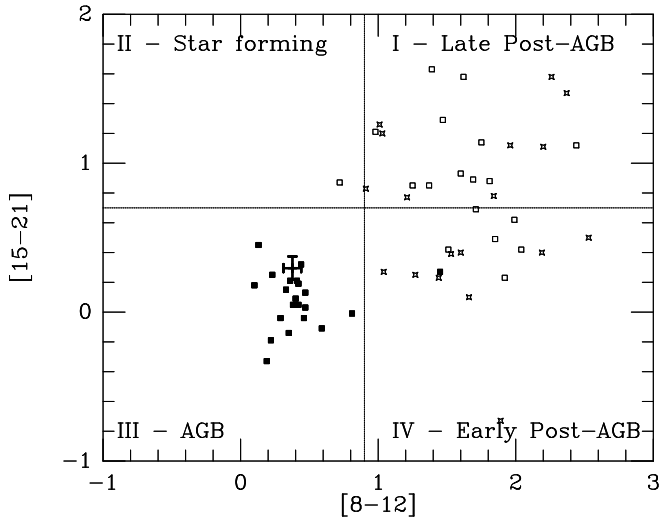}
\caption{Sevenster's suggested 4-quadrant MSX colour-colour diagram in which quadrants III 
and I/IV respectively separate OH masers into LI (filled squares) and RI objects (open squares).  
OH\,354.88-0.54 is the large cross among the LI sources, showing its 
1$\sigma$ uncertainties in the MSX colours. Open star symbols represent sources not yet 
definitively assigned to RI status.}
\label{msmsx}
\end{figure}

Figure~\ref{msiras} is an adaptation of Sevenster's {\it IRAS} colour-colour plane showing 
likely AGB and post-AGB stars currently under investigation at ATNF (R. Deacon, 
priv. communication).  Deacon's sample of 85 stars includes some objects chosen by 
their IRAS colours, and others by their MSX colours. 
We calculated the three individual sets of {\it IRAS} colours 
from the WSDB time-resolved measurements of OH\,354.88-0.54 and plotted those in Figure~\ref{msiras}. 
Within the photometric uncertainties, all three locations, corresponding to IR light curve phases 
0.57-0.69 during the {\it IRAS} mission, lie along the evolutionary track of van de Veen \& Habing (1988),
moving monotonically up the track with time (large crosses).  The {\it IRAS} colours of OH\,354.88-0.54 
place it near the lower right corner of the LI zone, and its expansion
speed of 14.9~km s$^{-1}$ also suggests its proper association with the LI sources.
Figure~\ref{msmsx} is likewise an adaptation of Sevenster's suggested MSX colour-colour plane for
distinguishing LI from RI sources.  OH\,354.88-0.54 lies with 
the LI sources. Table~\ref{tphot} summarizes median or mean IR photometry for OH\,354.88-0.54.
The resulting bolometric luminosity from this table is 3.1$\times$10$^4$~L$_\odot$.
Therefore, all this object's MIR attributes are consistent with a somewhat massive and still 
pulsating variable with the properties of the LI stars.  

\begin{table}
\begin{center}
\caption{Median NIR and Mean MIR/FIR Photometry from the Literature of OH\,354.88-0.54}
\label{tphot}
\begin{tabular}{lc}
Waveband & Mag/Flux Density\\ 
\hline
$J$                    & $ 14.3^m $  \\
$H$                    & $ 12.4^m $ \\
$K$                    & $ 11.6^m $ \\
$L^\prime$             & $ 3.5^m  $ \\
$M$                    & $ 1.1^m $ \\
{\it MSX} F(4$\mu$m) (Jy)            & $ 13.1\pm1.5 $  \\
{\it MSX} F(8$\mu$m) (Jy)            & $ 70.8\pm3.5 $  \\
{\it MSX} F(12$\mu$m) (Jy)           & $ 100.1\pm3.0 $    \\
{\it MSX} F(15$\mu$m) (Jy)           & $ 168.2\pm6.7 $    \\
{\it MSX} F(21$\mu$m) (Jy)           & $ 221.0\pm13.3$     \\
{\it IRAS} F(12$\mu$m) (Jy)           & $  104.0 $    \\
{\it IRAS} F(25$\mu$m) (Jy)           & $  291.5 $    \\
{\it IRAS} F(60$\mu$m) (Jy)            & $  234.5 $    \\
{\it IRAS} F(100$\mu$m) (Jy)           & $<736 $   \\
\hline
\end{tabular}
\end{center}
\end{table}

\section{The optical spectrum of the nebular ring}
\begin{figure}
\vspace{6cm}
\includegraphics{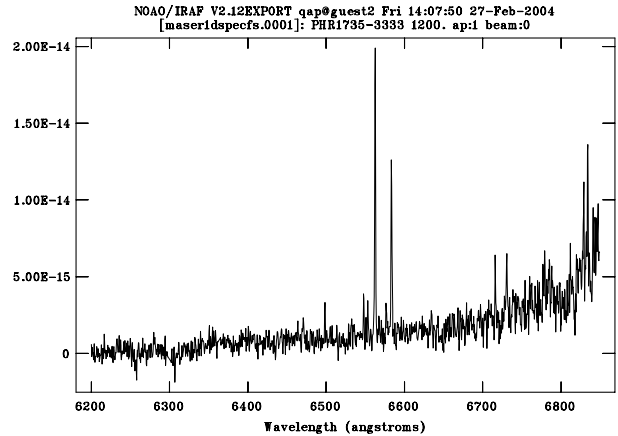}
\caption{The emission-line spectrum of the H$\alpha$ ring around OH\,354.88-0.54}
\label{nebspec}
\end{figure}

During the programme of spectroscopic follow-up of MASH PNe candidates
on the SAAO 1.9m telescope a long-slit (2$^\prime$) spectrum of the  
newly identified H$\alpha$ ring (within which the maser source is centred) was
taken on 2003 June 28 (JD2452819, phase 0.30). The instrumental configuration was similar to
that described by Parker \& Morgan (2003)  with the use of a high
dispersion 1200R grating giving $\sim0.4$\AA\ pixel$^{-1}$ over the range
6152-6912\AA. Wavelength calibration was via Copper-Argon arc lamp   
exposures taken on either side of the 1200s exposure on the H$\alpha$ ring.
The location of the slit is indicated in Figure~\ref{ring}. A nebular 
detection was obtained confirming the veracity of the optical shell.
The nebula spectrum was flux calibrated using the spectrophotometric  
standard star LTT6248 taken 15~min prior to the target exposure.
The spectrum is given in Figure~\ref{nebspec}, which shows H$\alpha$/[N{\sc ii}] and
[S{\sc ii}] emission lines together with evidence of a faint red continuum.  The abruptly
rising red continuum we attribute to contamination by a faint red star within the slit.  
Consequently, we regard the spectrum as that of a planetary 
nebula, because it is circularly symmetric, centred on an evolved exciting star of intermediate 
mass, still a red AGB star, and exhibits the canonical emisson lines associated with such nebulae.

The weighted average of the heliocentric radial velocities of these five lines is 
$-13\pm5$~km~s$^{-1}$,
equivalent to an LSR velocity of $-10\pm5$~km~s$^{-1}$.  The ratio of the sulphur lines, 
I(6717)/I(6731), is 0.59.  Adopting T$_e$=10$^4$~K, and using Table 5.7 of Osterbrock (1989)
yields N$_e$$\sim$4000~cm$^{-3}$.

One can estimate the ionized gas mass (M$_i$) in the PN from the correlation for Galactic 
disk PNe between M$_i$ and absolute size derived by Boffi \& Stanghellini (1994: their eqn.(9)).
We find M$_i$~=~0.35~M$_\odot$.  Using the expression for M$_i$ in a PN given by Boffi 
\& Stanghellini (1994) in their eqn.(5), treating the PN as a sphere of radius $\sim$10$^{18}$~cm
($\sim$20$^{\prime\prime}$ at 3.2~kpc), with N$_e$ determined
from the red [S{\sc ii}] lines, the deduced filling factor is 0.02.  Thus this PN is
characterized by a somewhat larger than average ionized mass and a smaller than average
filling factor.

\section{Conclusions}
We have identified a faint, circular nebula with a typical PN spectrum, centred on an extremely red,
long-period variable star associated with a bright OH 1612-MHz maser.  The PN spectrum is proof of the
reality and nature of the H$\alpha$ nebula, and the good centring of the OH maser suggests a causal link
between the long period variable and the nebula.  
The H$\alpha$ survey image covered the position of the OH/IR star on JD2451338 (1999 August 6)
at a light curve phase of the star's IR variation of 0.34, almost the same as the phase at which
the optical spectrum was obtained (0.30). 
The optical spectrum of the nebula samples the expanding gas only near the tangent
to the ring, giving us rather limited information on the outflow velocity of the gas 
far from the star.  The formal difference between the star's systemic velocity and that
of the nebular emission lines is $\sim$20~km~s$^{-1}$, but this must be a lower bound
on the true outflow speed because of our viewing geometry at the edge of the ring.
Could it be fast enough to collisionally ionize the ambient gas?  
The spectrum reveals no hint of [O{\sc i}] 6300/6363\AA\ emission lines, but our 
wavelength region did not cover [O{\sc ii}] 7320/7330\AA\ nor He{\sc i} 5876\AA, lines that could
signify gas cooling after a shock at a velocity between a few tens and $\sim$100 km~s$^{-1}$.
At much lower velocities, the resulting fractional ionization of H would be too low for us to detect.
If the nebular emission were due solely to shock interaction with the surrounding interstellar medium 
(ISM), one would not expect a uniformly bright nebula because the interaction would produce 
emission line cooling strongly dependent on local density inhomogeneities encountered in the ISM.
We plan to reobserve this nebula with a larger telescope, to widen our wavelength coverage, and to secure
new nebular spectra at several light curve phases.  

The central star appears to come from a progenitor of intermediate mass ($\geq4$~M$_\odot$).
It is still pulsating strongly, and 
moving along a suggested evolutionary track characteristic of Miras and OH/IR stars.  If OH\,354.88-0.54
were truly an LI source, Sevenster has argued that such objects are more likely to become bipolar PNe.
The high signal-to-noise OH spectral profile we present shows no hint of bipolarity
as yet, based upon the OH profiles predicted by Zijlstra et al. (2001) that signal the onset
of bipolarity in early post-AGB stars. The OH envelope
around this star is elongated in an E-W direction but sparser in maser emission at its poles (N-S), like 
an edge-on torus.  MIR stellar emission is elongated orthogonal to the OH maser emission, suggesting that
any outflowing gas in which dust grains have condensed is dominantly in the polar directions.  Therefore,
the MIR emission is likely to represent a more recent phenomenon than the accumulated toroid of OH gas.
Might this signify the earliest onset of the fast wind (Kwok, Purton \& Fitzgerald 1978) in this source?

The existence of the ionized nebula poses the question of what mechanism for ionization such an AGB star has
available to it.  The OH spectral profile indicates no evidence for a hot binary companion, precluding
direct radiative ionization of the circumstellar gas.  OH envelopes around long period variables are known 
to be sparsely populated and dominated by a relatively small number of clumps (Welty et al. 1987; van 
Langevelde et al. 1990).  Therefore, conceptually, a fast wind might penetrate through the toroid containing 
the OH molecules and far into the circumstellar region, producing collisional ionization of the accumulated 
material previously shed by the star during its AGB evolution.

One other scenario is worth considering as a possible source of nebular ionization.
If the central AGB star were a ``born-again" (e.g., Iben 1984) object that had undergone
a ``very late thermal pulse'' (``VLTP"; e.g., Bl\"{o}cker \& Sch\"{o}nberner 1997) as it entered the 
cooling track, then material would
have been shed and ionized by the hot central star as it created a PN at the end of 
its former AGB life.  Given sufficient remaining core helium, this late helium flash would have
returned the central star to the AGB within a very short period of time (a few hundred
years in a star initially of roughly solar mass and abundances 
(Lawlor \& MacDonald (2003), their Table~2)).  
Perhaps the currently OH-masing, long-period IR variable never had to ionize the nebulous ring. 
Empirical estimates for the luminosity of the star range from 1.7$\times$10$^4$ (Le Sidaner \&
Le Bertre 1996) to 3.1$\times$10$^4$~L$_\odot$ (Table 2, this paper), implying a bolometric magnitude
between -5.85 and -6.50, respectively.  Vassiliadis \& Wood (1993) have modeled the evolution of
AGB stars.  From their Table 2 one derives a minimum initial mass for the progenitor of OH\,354.88-0.54
of 4-5~$M\odot$ and an expected core mass of $\sim$0.8-0.9~$M\odot$.  This initial mass is higher than
those of the roughly solar-mass stars undergoing VLTPs discussed by Lawlor \& MacDonald (2003), and
Vassiliadis \& Wood (1994) did not incorporate VLTPs into their work on
the evolution of intermediate mass stars.  If a 4-5~$M\odot$ progenitor could undergo the VLTP
phenomenon, is it likely that the PN ejected after the first AGB phase of evolution would have remained
ionized while the star rejoined the AGB?  The recombination time scale for the PN around OH\,354.88-0.54,
with T$_e$=10$^4$, would be $\approx$10$^5$/N$_e$ yr (e.g. Marten \& Szczerba 1997), or 25~yr.  
It would be remarkable if such a progenitor could accomplish its post-VLTP evolution and 
return to the AGB within only a few years.  It is also noteworthy that this star has
shown no hint of the rapid changes of behaviour so characteristic of other 
possible VLTP sources such as V605 Aql and FG Sge.

Therefore, we speculate that we may be witnessing a hitherto unobserved phase of 
PN evolution, in which a PN has only recently started to appear around a star 
that is unequivocally still in its AGB phase.

\section{Acknowledgments}
MC thanks NASA for supporting this work under its Long Term Space Astrophysics and Astrophysics 
Data Analysis programmes, through grants NAG5-7936 and NNG04GD43G with UC Berkeley.  
The AAO has undertaken the H$\alpha$ survey on behalf of the astronomical community.
We thank Jim Caswell for his ever reliable and well-documented dossier on the early OH survey;
Nicolas Epchtein and Terry Jones for helping us to understand the discordant NIR measurements
from February 1981; Rachel Deacon for providing us with a table of the {\it IRAS} and MSX 
flux densities for her samples of maser sources; and Falk Herwig for valuable discussions
about the VLTP phenomenon in AGB stars.

\end{document}